\documentclass[aps,pra,reprint,amsmath,amssymb,showpacs,floatfix]{revtex4-1}

\usepackage{graphicx,units}

\newcommand{\ket}[1]{\ensuremath{\vert{#1\rangle}}} 
\newcommand{\bra}[1]{\ensuremath{{\langle #1}\vert}}
\newcommand{\braket}[2]{\ensuremath{{\langle #1}\vert{#2 \rangle}}}
\newcommand{\ketbra}[2]{\ensuremath{|{#1 \rangle}{\langle #2}|}}
\newcommand{\op}[1]{\hat{#1}}
\newcommand{\D}{\text{d}}
\newcommand{\I}{\text{i}}
\newcommand{\E}{\text{e}}
\providecommand{\abs}[1]{\left\lvert#1\right\rvert}
\newcommand{\bvec}[1]{\ensuremath{\mathbf{#1}}}

\newcommand{\bopvec}[1]{\ensuremath{\mathbf{\op{#1}}}}
\newcommand{\bopvecgr}[1]{\ensuremath{\mathbf{\op{\boldsymbol #1}}}}
\newcommand{\boldnabla}{\mbox{\boldmath$\nabla$}}

\usepackage[colorlinks,citecolor=blue,urlcolor=blue,linkcolor=blue,hyperindex,breaklinks]{hyperref}

\begin{document}

\title{Analysis of quantum-state disturbance in a \\protective measurement of a spin-1/2 particle}
\author{Maximilian Schlosshauer}
\affiliation{\small Department of Physics, University of Portland, 5000 North Willamette Boulevard, Portland, Oregon 97203, USA}

\begin{abstract} 
A distinguishing feature of protective measurement is the possibility of obtaining information about expectation values while making the disturbance of the initial state arbitrarily small. Quantifying this state disturbance is of paramount importance. Here we derive exact and perturbative expressions for the state disturbance and the faithfulness of the measurement outcome in a model describing a spin-$\frac{1}{2}$ particle protectively measured by an inhomogeneous magnetic field. We determine constraints on the experimentally required field strengths from bounds imposed on the allowable state disturbance. We also demonstrate that the protective measurement may produce an incorrect result even when the particle's spin state is unaffected by the measurement, and show that successive measurements using multiple magnetic fields produce the same state disturbance as a single measurement involving a superposition of these fields. Our results supply comprehensive understanding of a paradigmatic model for protective measurement, may aid the experimental implementation of the measurement scheme, and illustrate fundamental properties of protective measurements. \\[-.1cm]

\noindent Journal reference: \emph{Phys.\ Rev.\ A\ }\textbf{92}, 062116 (2015), DOI: \href{http://dx.doi.org/10.1103/PhysRevA.92.062116}{10.1103/PhysRevA.92.062116}
\end{abstract}

\maketitle

\section{Introduction}

Any quantum measurement changes (``disturbs'') the quantum state of the measured system. Protective measurement \cite{Aharonov:1993:qa,Aharonov:1993:jm,Aharonov:1996:fp,Dass:1999:az,Vaidman:2009:po,Gao:2014:cu} is a scheme for measuring expectation values of a quantum system in a way that makes this state disturbance arbitrarily small. For such a measurement to obtain, the system is coupled weakly to the apparatus, and the initial state of the system is required to be in a nondegenerate eigenstate of its Hamiltonian. Neither the initial state nor the Hamiltonian need to be known. Besides being an important and unusual instance of a quantum measurement, protective measurement offers the possibility of successively carrying out measurements of expectation values of different observables while the system is likely to remain in its initial quantum state. This state may then be reconstructed from the measured expectation values with in principle arbitrarily high fidelity \cite{Aharonov:1993:qa,Aharonov:1993:jm,Aharonov:1996:fp,Dass:1999:az}. In this way, protective measurement may provide a route toward state tomography of single quantum systems. 

Two essential aspects of protective measurement remain insufficiently studied. The first is to actually quantify the state disturbance, rather than to simply consider the idealization of an infinitely weak and infinitely long measurement interaction, for which no state disturbance occurs. The second is to study concrete models for implementing protective measurements. In particular, what has not yet been done is to study the issue of state disturbance in the context of such a model, so one can better understand the physics and parameter choices that would need to go into an experiment realizing a protective measurement with minimal state disturbance.
Our paper addresses this open problem. 

Specifically, we consider the protective measurement of the state of a spin-$\frac{1}{2}$ particle by an inhomogeneous magnetic field using a Stern--Gerlach-like setup. This paradigmatic and experimentally relevant model was first studied in Refs.~\cite{Aharonov:1993:qa, Dass:1999:az}, but those studies merely demonstrated one basic feature of protective measurement, namely, how information about the expectation value becomes encoded in the shift of the apparatus pointer. The studies only considered in the limit of an infinitesimally weak inhomogeneous magnetic field applied for an infinite amount of time (and infinitely rapidly turned on and off), without attending to the crucial issue of state disturbance. Here, we revisit this model but develop a significantly more general account of it that allows us to precisely quantify the amount of state disturbance as a function of the relevant physical parameters. Given the importance of two-level systems (qubits) in quantum mechanics and quantum information processing, our model is of great theoretical and practical interest. Because the model is exactly solvable, we can also use it as a tool for assessing the accuracy of approximate and perturbative solutions that are needed to treat most other models of protective measurement. By applying a recently developed perturbative approach \cite{Schlosshauer:2014:pm}, we are also able to explore how time-dependent couplings between system and apparatus help reduce the state disturbance. 

The main result of our paper is the derivation of both exact and approximate expressions for the state disturbance predicted by the model. This has direct physical implications, because it allows us to estimate the values of the magnetic-field parameters that would be required to implement a protective measurement that has a suitably low probability of disturbing the initial state. We also show that multiple simultaneous protective measurements do not result in smaller cumulative state disturbance when compared to a successive implementation of these measurements. More generally, our analysis identifies and illustrates properties of protective measurements and of quantum measurements in general, such as complementarity and the tradeoff between information gain and disturbance. 

This paper is organized as follows. In Sec.~\ref{sec:model-prot-meas} we describe the model for the protective quantum measurement of a spin-$\frac{1}{2}$ particle. In Sec.~\ref{sec:state-disturbance} we derive exact and perturbative solutions for the state disturbance incurred during the protective measurement. In Sec.~\ref{sec:quant-state-reconstr}, we compare the state disturbance for successive and simultaneous protective measurements. In Sec.~\ref{sec:pointer-shift} we explore a hitherto overlooked issue, namely, the possibility of a protective measurement's failing due to a reversed momentum shift. We discuss our results in Sec.~\ref{sec:disc-concl}.

\section{\label{sec:model-prot-meas}Measurement model}

In a protective measurement \cite{Aharonov:1993:qa,Aharonov:1993:jm,Aharonov:1996:fp,Dass:1999:az,Vaidman:2009:po}, the quantum system interacts weakly with a measuring apparatus via an interaction Hamiltonian given by
\begin{equation}\label{eq:lalaa}
\op{H}_\text{int}(t) = g(t)\op{O} \otimes \op{P}.
\end{equation}
Here $\op{O}$ is the observable protectively measured on the system, and $\op{P}$ is the operator that generates a corresponding shift of the pointer of the apparatus. The coupling function $g(t)$ describes the time dependence of the strength of the system--apparatus interaction and is normalized according to
\begin{equation}\label{eq:lalaanorm}
\int_{0}^{T} \D t\, g(t) =1,
\end{equation}
where $t=0$ marks the onset of the protective measurement and $T$ is the total measurement time [thus $g(t)=0$ for $t<0$ and $t>T$]. It follows that the average coupling strength over the course of the measurement is equal to $1/T$ for all choices of $g(t)$. The normalization condition \eqref{eq:lalaanorm} not only imposes an inversely proportional relationship between this average coupling strength and the measurement time $T$, but also ensures that the total pointer shift produced by the protective measurement is independent of the particular functional form of $g(t)$ \cite{Schlosshauer:2014:pm}. 

Consider now the following model of a protective measurement of a spin-$\frac{1}{2}$ particle by a magnetic field, first discussed in Ref.~\cite{Aharonov:1993:jm} (see also Sec.~II\,E of Ref.~\cite{Dass:1999:az}). A spin-$\frac{1}{2}$ particle travels through a uniform magnetic field of unknown direction and magnitude. The field provides the protection of the particle's initial spin state, which is quantized along the direction of the field. Information about expectation values of spin components along different axes is obtained by introducing weak inhomogeneous magnetic fields in different directions, which produce a change in the particle's momentum (resulting in a displacement of its trajectory). The expectation values of different spin components can then be obtained from measuring these momentum shifts, and the spin state of the particle may be reconstructed from the measured expectation values. 

Denoting the uniform magnetic field by $\bvec{B}_0=B_0\bvec{e}_z$, the spin part of the Hamiltonian of the particle is
\begin{equation}\label{eq:vshvbjfdjhvs}
\op{H}_S = -  \mu \bopvecgr{\sigma} \cdot \bvec{B}_0 = -\mu B_0 \op{\sigma}_z,
\end{equation}
where $\mu$ is the magnetic moment of the particle \footnote{It is important to emphasize that although we have specified the direction of the field $\bvec{B}_0$ to enable the subsequent mathematical description of the model, in an actual realization of the protective measurement the field's direction and magnitude will be \emph{a priori} unknown (see also the discussion in Refs.~\cite{Aharonov:1993:jm} and \cite{Dass:1999:az}). Indeed, an observer who knows $\bvec{B}_0$ would also know $\op{H}_S$ and could therefore perform a simple projective measurement in the eigenbasis of $\op{H}_S$ to determine the spin state, eliminating the need to perform a protective measurement.}. The eigenstates of $\op{H}_S$ are the eigenstates $\ket{\pm}$ of $\op{\sigma}_z$, with eigenvalues $E_\pm=\mp \mu B_0$ and corresponding transition frequency
\begin{equation}\label{eq:vshvbjs}
\omega_0 = \frac{2\mu B_0}{\hbar}. 
\end{equation}
The particle is assumed to be in one of these two eigenstates. Between $t=0$ and $t=T$, the particle traverses a region in which an additional inhomogeneous time-dependent magnetic field 
\begin{equation}\label{eq:measfield}
\bvec{B}_1(\bvec{x},t) = g(t) \beta q  \bvec{n},
\end{equation}
is present, where $\bvec{n}$ is the (known) direction, $g(t)$ is the time dependence of the field strength, and $q$ is the position coordinate in the direction of $\bvec{n}$ \footnote{As already noted in Ref.~\cite{Aharonov:1993:jm}, since Eq.~\eqref{eq:measfield} has nonzero divergence, it violates Maxwell's equations and cannot represent a real physical magnetic field. However, a suitable divergence-free inhomogeneous field is easily constructed \cite{Anandan:1993:uu}. Such a field effectively acts as a superposition of three fields of the kind given by Eq.~\eqref{eq:measfield} and leads to the same cumulative momentum shift and state disturbance (see also Sec.~\ref{sec:quant-state-reconstr}). Without loss of generality, we may therefore restrict our attention to the field defined by Eq.~\eqref{eq:measfield}.}.  We refer to $\bvec{B}_0$ as the protection field and $\bvec{B}_1$ as the measurement field. The interaction Hamiltonian is taken to be
\begin{equation}\label{eq:1dvhjbbdhvbdhjv}
\op{H}_\text{int}(\bvec{x}, t) = -   \mu \bopvecgr{\sigma} \cdot \bvec{B}_1(\bvec{x},t) = -g(t) \mu\beta (\bopvecgr{\sigma} \cdot
  \bvec{n}) \otimes \op{q}.
\end{equation}
Comparison with Eq.~\eqref{eq:lalaa} shows that the system observable to be protectively measured is the spin component $\bopvecgr{\sigma} \cdot \bvec{n}$ in the direction $\bvec{n}$ of the measurement field, while $\op{q}$ is the apparatus observable that couples to the spin component. The apparatus observable $\op{q}$ does not commute with the self-Hamiltonian $\op{H}_A =\bopvec{p}^2/2m$ of the apparatus associated with the phase-space degree of freedom of the particle, and therefore $\op{q}$ is not a constant of motion. Because this noncommutativity complicates the mathematical treatment while leaving unaffected the possibility or the physics of a protective measurement \cite{Dass:1999:az}, Refs.~\cite{Aharonov:1993:jm,Dass:1999:az} have considered the particle in its rest frame, such that $\op{H}_A =0$. Adopting this approach, the state of the particle for $t<0$ is
\begin{equation}\label{eq:b333hjsvbhsaasas}
\Psi(\bvec{x},t) = \ket{\pm}\exp \left( \pm \frac{\I \mu B_0 t}{\hbar} \right) = \ket{\pm}\exp \left( \pm \frac{\I \omega_0 t}{2} \right),
\end{equation}
and the total Hamiltonian $\op{H}(t)$ defining our model is 
\begin{align}\label{eq:bhjsvbhsaasas}
\op{H}(\bvec{x}, t)  &= \op{H}_S + \op{H}_\text{int}(\bvec{x}, t) \notag \\ &= -\mu B_0 \op{\sigma}_z -g(t) \mu\beta (\bopvecgr{\sigma} \cdot
  \bvec{n}) \otimes \op{q}.
\end{align}

\section{\label{sec:state-disturbance}State disturbance}

\subsection{Constant coupling}

We now derive our main result, an expression for the state disturbance of the initial spin state by the protective measurement. We first consider the time-independent coupling function (hereafter ``constant coupling'') defined by
\begin{equation}\label{eq:lbivdddhv}
g(t) = \begin{cases} 1/T, & 0 \le t \le T, \\ 0, & \text{otherwise}. \end{cases}
\end{equation}
Then the Hamiltonian~\eqref{eq:bhjsvbhsaasas} is time-independent,
\begin{equation}\label{eq:vihdgsv}
\op{H}(\bvec{x})  =  -  \mu\bopvecgr{\sigma} \cdot \bvec{B}(\bvec{x}),
\end{equation}
where
\begin{equation}\label{eq:vihdgsv22}
\bvec{B}(\bvec{x}) = B_0 \bvec{e}_z + \frac{1}{T} \beta q \bvec{n},
\end{equation}
which shows that the strength of the measurement field scales as $1/T$. The eigenvectors of the Hamiltonian~\eqref{eq:vihdgsv} are
\begin{subequations}\label{eq:huhuhu}
\begin{align}
\ket{+}_\bvec{x} &= \cos\frac{\theta(\bvec{x})}{2}\ket{+} + \sin\frac{\theta (\bvec{x})}{2}\E^{\I \phi (\bvec{x})}\ket{-}, \label{eq:huhuhu1}\\ 
\ket{-}_\bvec{x} &= \sin\frac{\theta(\bvec{x})}{2}\ket{+} - \cos\frac{\theta (\bvec{x})}{2}\E^{\I \phi (\bvec{x})}\ket{-},\label{eq:huhuhu2}
\end{align}
\end{subequations}
with corresponding eigenvalues $E_\pm(\bvec{x})=\mp \mu B(\bvec{x})$. Here $\theta (\bvec{x})$ and $\phi (\bvec{x})$ are the polar and azimuthal angles of $\bvec{B}(\bvec{x})$. Note that $\theta (\bvec{x})$ is also the angle between $\bvec{B}(\bvec{x})$ and $\bvec{B}_0$, and that $\phi (\bvec{x})$ is equal to the (fixed) azimuthal angle $\eta$ of the field-direction vector $\bvec{n}$. If the initial spin state is $\ket{+}= \cos\frac{\theta(\bvec{x})}{2}\ket{+} _\bvec{x}  + \sin\frac{\theta(\bvec{x})}{2}\ket{-}_\bvec{x}$, then at $t=T$ it is 
\begin{align}\label{eq:vihdgs7cf6gv}
\ket{\psi(\bvec{x}, T)} &= \cos\frac{\theta(\bvec{x})}{2} \exp\left( \frac{\I \mu B(\bvec{x}) T}{\hbar}\right) \ket{+} _\bvec{x}  \notag \\ &\quad + \sin\frac{\theta(\bvec{x})}{2} \exp\left(- \frac{\I \mu B(\bvec{x}) T}{\hbar}\right) \ket{-}_\bvec{x}.
\end{align}
Existing treatments of the model \cite{Aharonov:1993:jm,Dass:1999:az} have only considered the limit of (infinitely) large measurement times $T$, such that $\bvec{B}_1 \ll \bvec{B}_0$. Then $\theta (\bvec{x}) \ll 1$, and thus
\begin{equation}\label{eq:hbvdvbhj}
B(\bvec{x}) \approx \bvec{B}(\bvec{x}) \cdot \bvec{e}_z = B_0 + \frac{1}{T} \beta q \cos\gamma,
\end{equation}
where $\gamma$ is the polar angle of $\bvec{n}$, which is also the angle between $\bvec{B}_1(\bvec{x})$ and $\bvec{B}_0$. In this limit, the state~\eqref{eq:vihdgs7cf6gv} becomes
\begin{equation}
\ket{\psi(\bvec{x}, T)} \approx \exp\left( \frac{\I \omega_0 T}{2} \right) \exp\left(\frac{\I \mu \beta q  \cos\gamma }{\hbar} \right) \ket{+} .
\end{equation}
This is the familiar result of protective measurement originally derived in Ref.~\cite{Aharonov:1993:jm}. The system remains, with arbitrarily large probability, in its initial state $\ket{+}$, while the term $\exp\left( \I \mu \beta q  \cos\gamma /\hbar\right)$ induces a change in momentum (pointer shift) in the direction of $\bvec{n}$ of size $\Delta p = \mu\beta\cos\gamma$. Since $\cos\gamma = \bra{+} \bopvecgr{\sigma} \cdot \bvec{n}  \ket{+}$,
the momentum shift can be written as $\Delta \bvec{p} = \mu\beta  \bra{+} \bopvecgr{\sigma} \cdot \bvec{n}  \ket{+} \bvec{n}$, which is proportional to the expectation value of the system observable $\bopvecgr{\sigma} \cdot \bvec{n}$ in the initial state $\ket{+}$. 

State disturbance manifests itself in a nonzero probability amplitude for finding the system in the state $\ket{-}$ at $t=T$. We write the final state as
\begin{align}\label{eq:1fbhjsbfk4554jaa}
\ket{\Psi(\bvec{x},T)} &= A_+(\bvec{x},T)\ket{+} + A_-(\bvec{x},T) \ket{-},
\end{align}
where the phase-space part has been absorbed into $A_\pm(\bvec{x},T) = \braket{\pm}{\psi(\bvec{x}, T)} $. The amplitude $A_-(\bvec{x},T)$ is the probability amplitude for the transition $\ket{+} \rightarrow \ket{-}$, and we therefore take $\abs{A_-(\bvec{x},T)}^2$ as a measure for the state disturbance. From Eqs.~\eqref{eq:huhuhu} and \eqref{eq:vihdgs7cf6gv}, we find
\begin{equation}\label{eq:sy7syvvfvihdgs7cf6gv}
A_-(\bvec{x},T)= \I \E^{\I \eta} \sin \theta(\bvec{x}) \sin \left( \frac{\mu B(\bvec{x}) T}{\hbar}\right).
\end{equation}
Using Eq.~\eqref{eq:vihdgsv22}, we have
\begin{align}\label{eq:vdh5478457}
B(\bvec{x}) &=\sqrt{ \bvec{B}(\bvec{x}) \cdot \bvec{B}(\bvec{x}) } =B_0 \sqrt{1 + \xi^2 + 2 \xi \cos\gamma},
\end{align}
where 
\begin{equation}\label{eq:xi}
\xi=\xi(q,T)=\frac{\beta q}{B_0T}
\end{equation}
measures the relative field strength $B_1/B_0$, where $B_1$ has been evaluated at position $q$ along direction $\bvec{n}$. We may associate $q$ with the measured location of the particle when it has completed its passage through the measurement field. 

Explicitly evaluating the term $\sin \theta(\bvec{x})$ appearing in  Eq.~\eqref{eq:sy7syvvfvihdgs7cf6gv},
\begin{align}\label{eq:vbdiubdfiubdfisaks}
\sin \theta (\bvec{x}) &= \frac{\sqrt{ \left[\bvec{B}(\bvec{x}) \cdot \bvec{e}_x\right]^2 +  \left[\bvec{B}(\bvec{x}) \cdot \bvec{e}_y\right]^2}}{B(\bvec{x})} \notag\\&= \frac{\xi \sin\gamma}{\sqrt{ 1 + \xi^2 + 2 \xi \cos\gamma}},
\end{align}
the transition amplitude can be written as
\begin{multline}\label{eq:sy7syvvfvihdgs7cf6gv0}
A_-(\bvec{x},T) \equiv A_-(q,\gamma,T) \\ = \I \E^{\I \eta}\frac{\mu\beta q}{\hbar} \sin\gamma\,\mathrm{sinc} \left( \frac{\omega_0 T }{2} \sqrt{1 + \xi^2 + 2 \xi \cos\gamma}\right),
\end{multline}
where $\mathrm{sinc}(x)=\sin(x)/x$. Note that the quantity $\xi$ appearing in Eq.~\eqref{eq:sy7syvvfvihdgs7cf6gv0} depends on $T$, $B_0$, $\beta$, and $q$ [see Eq.~\eqref{eq:xi}].  We have also explicitly added the variable $\gamma$ to the argument to indicate that the transition amplitude depends on this angle approximately as $\sin^2 \gamma$. 

Note that Eq.~\eqref{eq:sy7syvvfvihdgs7cf6gv0} is zero if the sinc function is equal to zero. However, we cannot deliberately target these zeros to evade state disturbance, since it would require precise \emph{a priori} knowledge of $B_0$ and $\gamma$ that is unavailable in a protective measurement. We may therefore disregard the oscillations of $\mathrm{sinc}(x)$ and replace it by its decay envelope given by $1/x$ for $x \gtrsim 1$, such that
\begin{equation}\label{eq:sydvhiuvhuis7cf6gv}
A_-(\xi,\gamma)= \I \E^{\I \eta} \sin\gamma\frac{\xi}{\sqrt{1 + \xi^2 + 2 \xi \cos\gamma}},
\end{equation}
and thus the transition probability is
\begin{equation}\label{eq:sydvhiuvhuii90hj232}
P_-(\xi,\gamma) = \abs{A_-(T)}^2 =  \frac{\xi^2 \sin^2\gamma}{1 + \xi^2 + 2 \xi \cos\gamma}.
\end{equation}
We have written the argument as $(\xi,\gamma)$ to emphasize that the amount of state disturbance depends only on the relative strength $\xi$ of the measurement field (which is inversely proportional to $T$) and the angle $\gamma$ (which is determined by the measured observable $\bopvecgr{\sigma} \cdot \bvec{n}$).

\begin{figure}
\includegraphics[scale=0.9]{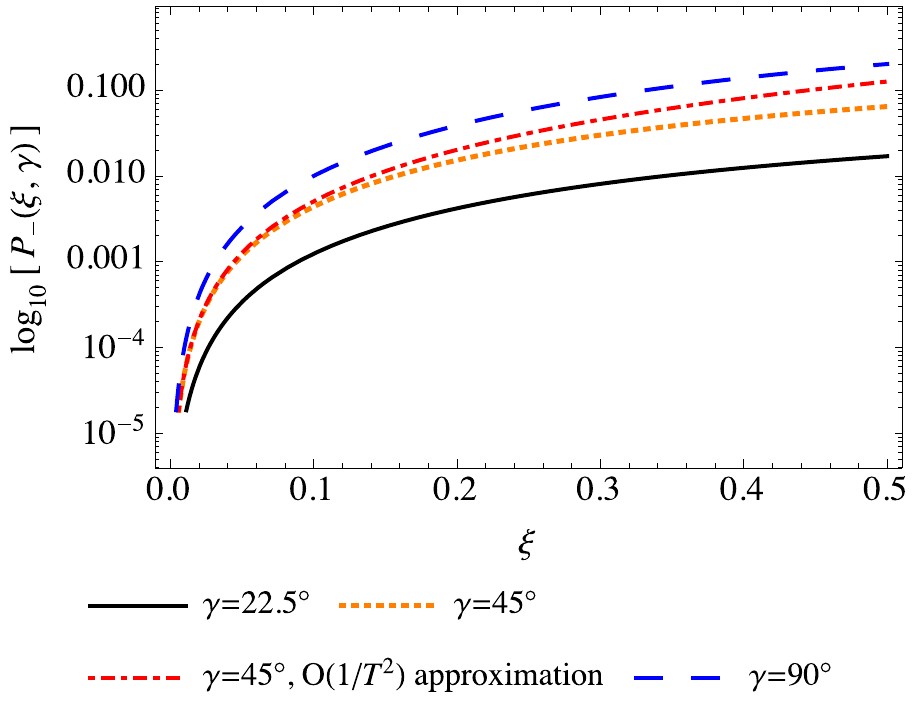}
\caption{\label{fig:dist}(Color online) Amount of state disturbance introduced during the protective measurement as quantified by the probability $P_-(\xi,\gamma)$ of transitioning to the orthogonal state, shown for constant coupling as a function of the dimensionless parameter $\xi=\beta q/B_0T$ for three different values of $\gamma$. The parameter $\xi$ measures the strength of the measurement field $\bvec{B}_1(\bvec{x})$ relative to the strength of the protection field $\bvec{B}_0$, and $\gamma$ represents the angle between $\bvec{B}_1(\bvec{x})$ and $\bvec{B}_0$. In addition to the probabilities calculated from the exact expression \eqref{eq:sydvhiuvhuii90hj232}, we also show the $O(1/T^2)$  approximation obtained from Eq.~\eqref{eq:bgb11j} for $\gamma=45^\circ$.}
\end{figure}
 
Figure~\ref{fig:dist} shows the state disturbance calculated from the exact expression~\eqref{eq:sydvhiuvhuii90hj232} as a function of $\xi$ for $\gamma=22.5^\circ$, $45^\circ$, and $90^\circ$. The reduction of the state disturbance achieved by decreasing $\xi$ is clearly seen, where a decrease in $\xi$ corresponds to an increase in the measurement time $T$ with its inversely proportional effect on the strength of the measurement field [see Eq.~\eqref{eq:vihdgsv22}]. We also see how the state disturbance grows with $\gamma$ as the direction of the measured spin component moves further away from the direction of the protection field $\bvec{B}_0$. Specifically, in the worst-case scenario $\gamma=90^\circ$ for state disturbance, in order not to exceed a probability $P_\text{max}$ of state disturbance we need $\xi \le \sqrt{P_\text{max}/(1-P_\text{max})}$. It follows that $\xi=0.1$ is sufficient for remaining within a 1\% probability of transitioning to the orthogonal state, regardless of the particular value of $\gamma$ (making such thresholds hold for all possible values of $\gamma$ is important because in an experimental setting the value of $\gamma$ will not be known \emph{a priori}).

Assuming weak measurement ($\xi \ll 1$), we can Taylor-expand Eq.~\eqref{eq:sydvhiuvhuis7cf6gv} to first order in $\xi$,
\begin{align}\label{eq:bgb11jy78g}
A_-(\xi, \gamma)&= \I \E^{\I \eta} \xi \sin\gamma + O(\xi^2)\notag\\&= \I \E^{\I \eta} \frac{\beta q}{B_0 T} \sin\gamma + O(1/T^2).
\end{align}
To second order in $1/T$, the corresponding transition probability is therefore
\begin{equation}\label{eq:bgb11j}
P_-(\xi, \gamma) \approx \xi^2\sin^2\gamma = \left(\frac{\beta q}{B_0}\right)^2 \frac{\sin^2\gamma }{T^2}.
\end{equation}
This expression exhibits the $1/T^2$ dependence familiar from first-order perturbation theory \cite{Schlosshauer:2014:tp} and the sinusoidal dependence on $\gamma$. It is shown in Fig.~\ref{fig:dist} (dashed-dotted line) as a function of $\xi$ for $\gamma=45^\circ$. We see that it represents an excellent approximation to the exact transition amplitude given by Eq.~\eqref{eq:sydvhiuvhuii90hj232} for the case $\xi \ll 1$ relevant to protective measurement. For larger $\xi \lesssim 1$, it produces a small overestimate of the state disturbance.

Let us investigate the extent to which the weak-measurement condition of small $\xi$ may hold in a concrete experimental setting. We consider a Stern--Gerlach experiment based on evaporated potassium atoms ($\mu = \unit[9.3 \times 10^{-24}]{J/T}$) and a movable hot-wire detector, a common modern implementation \cite{Daybell:1967:sg}, and estimate the required inhomogeneity of the measurement field to achieve an appreciable displacement. The magnitude of the momentum shift is $\Delta p = \mu\beta  \cos\gamma$, and thus the force on the particle due to the measurement field is $F = \mu(\beta /T) \cos\gamma$, where $\beta/T$ corresponds to the measurement-field gradient $\abs{\boldnabla B_1}$. With a typical oven temperature $T_\text{oven}=\unit[500]{K}$, the most probable velocity of a potassium atom emitted from the oven is $v=\sqrt{2 k_BT_\text{oven}/m} \approx \unit[450]{m/s}$. Then the spatial displacement in the direction $\boldnabla B_1$ of the inhomogeneity is
\begin{equation}
\Delta s = \frac{\mu\beta \cos\gamma}{2m T} T^2 = \frac{\mu \abs{\boldnabla B_1}\cos\gamma}{2mv^2} d^2 = \frac{\mu \abs{\boldnabla B_1}\cos\gamma}{4 k_BT_\text{oven}} d^2.
\end{equation}
To achieve a displacement $\Delta s = \unit[0.5]{mm}$, the required measurement-field gradient (setting $\gamma =45^\circ$ from here on) is $\boldnabla B_1 \approx \unit[20]{T/m}$, which is a typical value in a Stern--Gerlach experiment. 

Experimentally achievable strengths of a continuous uniform magnetic field are around $\unit[10]{T}$. For $B_0=\unit[10]{T}$, the weak-measurement condition of small $\xi$, which is here given by $(\boldnabla B_1) d/B_0$, is reasonably well fulfilled, since $(\boldnabla B_1) d/B_0 \approx  0.2$. From Eq.~\eqref{eq:bgb11j}, this value implies a transition probability of 2\%. Improvement is possible by increasing the size $d$ of the measurement region, since $\Delta s \propto d^2$ while $\xi$ increases only linearly with $d$. For example, for $d=\unit[1]{m}$, the same displacement $\Delta s = \unit[0.5]{mm}$ requires only $\boldnabla B_1 \approx \unit[0.2]{T/m}$. We may then lower the uniform field to $B_0=\unit[1]{T}$ while maintaining the previous values for the field ratio $(\boldnabla B_1) d/B_0 \approx  0.2$ and the state disturbance. Alternatively, if we maintain the uniform field at $B_0=\unit[10]{T}$, the field ratio is reduced to 0.02, leading to a 100-fold reduction in state disturbance. 

One experimental challenge will be to supply a sufficiently strong uniform magnetic field with a small but well-defined inhomogeneity over an extended region in space. Furthermore, since the magnitude of the displacement depends on the atomic velocities, which follow a thermal distribution upon the emission of the atoms from the oven, resolving the $\cos\gamma$ dependence of the momentum shift will necessitate a selection stage that produces atoms with a narrow range of velocities and directions. Experimental challenges of this kind notwithstanding, our numerical estimates suggest that the protective measurement considered in this paper may be experimentally realizable, at least in principle, using a standard Stern--Gerlach experiment with a superposed strong uniform magnetic field of practically achievable strength.

\subsection{\label{sec:state-dist-time}State disturbance for time-dependent measurement fields}

Explicitly time-dependent couplings $g(t)$ between system and apparatus, such as those describing a gradual turn-on and turnoff of the measurement interaction, are of great relevance to protective measurement, since they allow for a significant reduction of the state disturbance compared to the case of constant coupling \cite{Schlosshauer:2014:pm}. Here we illustrate this reduction in the context of our model. We consider the interaction Hamiltonian~\eqref{eq:1dvhjbbdhvbdhjv} as a time-dependent perturbation to $\op{H}_S =-\mu B_0 \op{\sigma}_z$ and express the state-vector amplitudes $A_\pm(\bvec{x},T)$ appearing in Eq.~\eqref{eq:1fbhjsbfk4554jaa} as a perturbative series (Dyson series), $A_\pm(\bvec{x},T) = \sum_{\ell=0}^\infty A_\pm^{(\ell)}(\bvec{x},T)$.
Here $A^{(\ell)}_\pm(\bvec{x},T)$ is the expression for the $\ell$th-order correction to the zeroth-order amplitudes $A^{(0)}_+(\bvec{x},T)=1$ and $A^{(0)}_-(\bvec{x},T)=0$. For protective measurements, the first-order transition amplitude $A_\pm^{(1)}(\bvec{x},T)$ is a reliable measure of the state disturbance \cite{Schlosshauer:2014:pm}. Applying the formalism of Ref.~\cite{Schlosshauer:2014:pm} to our model, we find
\begin{align}\label{eq:8dhj7gr7ss82}
A_-^{(1)}(q,\gamma,T) &= \I \E^{-\I \omega_0 T/2} \frac{\mu\beta q}{\hbar} \,\E^{\I \eta} \sin\gamma \int_{0}^{T} \D t\, \E^{\I \omega_0 t} g(t).
\end{align}
For the case of constant coupling considered before, this becomes (again disregarding the oscillations of the sinc function)
\begin{align}\label{eq:bdghv7vgh7g87vgy}
A_-^{(1)}(q,\gamma,T) &= \I \frac{\mu\beta q}{\hbar} \sin\gamma \,\E^{\I \eta}\frac{1}{\omega_0 T/2} = \I \E^{\I \eta}\xi \sin\gamma,
\end{align}
which is the same as Eq.~\eqref{eq:bgb11jy78g}, the approximation of Eq.~\eqref{eq:sy7syvvfvihdgs7cf6gv0} for weak measurement fields. Indeed, by evaluating the higher-order corrections $A_-^{(\ell)} (q,\gamma,T)$, one finds that $A_-^{(\ell)} (q,\gamma,T)$ is equal to the term of order $\ell$ in $\xi$ in the Taylor expansion of the exact solution for $A_-(q,\gamma,T)$, Eq.~\eqref{eq:sy7syvvfvihdgs7cf6gv0}. This agreement can be explained by noting that the Dyson series is a power series in the perturbation-strength parameter, here represented by $\xi$. 

\begin{figure}

\begin{flushleft}
{\small (a)}
\end{flushleft}

\vspace{-.5cm}

\includegraphics[scale=0.85]{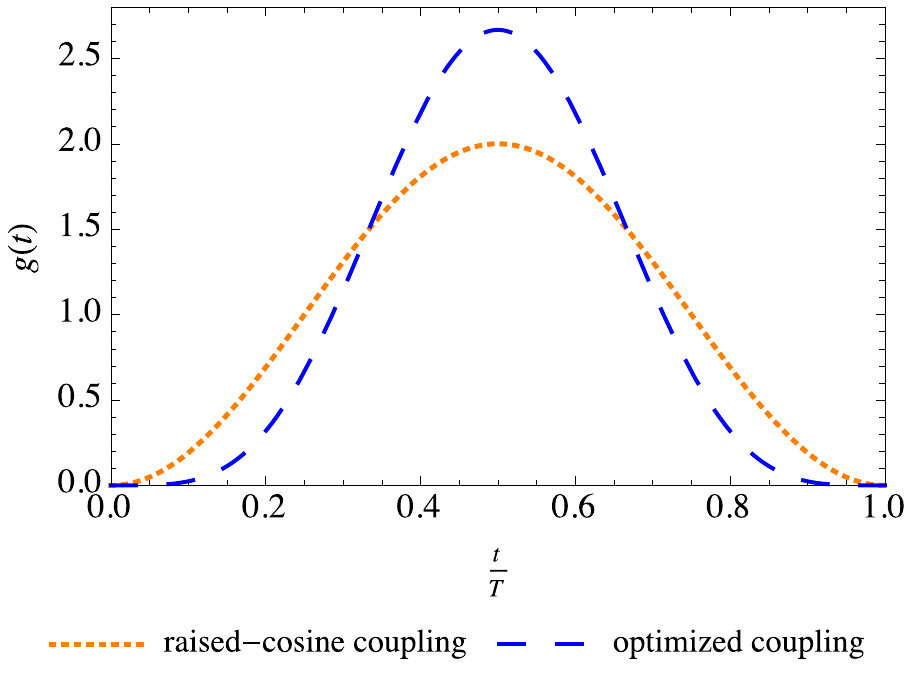}


\begin{flushleft}
{\small (b)}
\end{flushleft}

\vspace{-.5cm}

\includegraphics[scale=0.85]{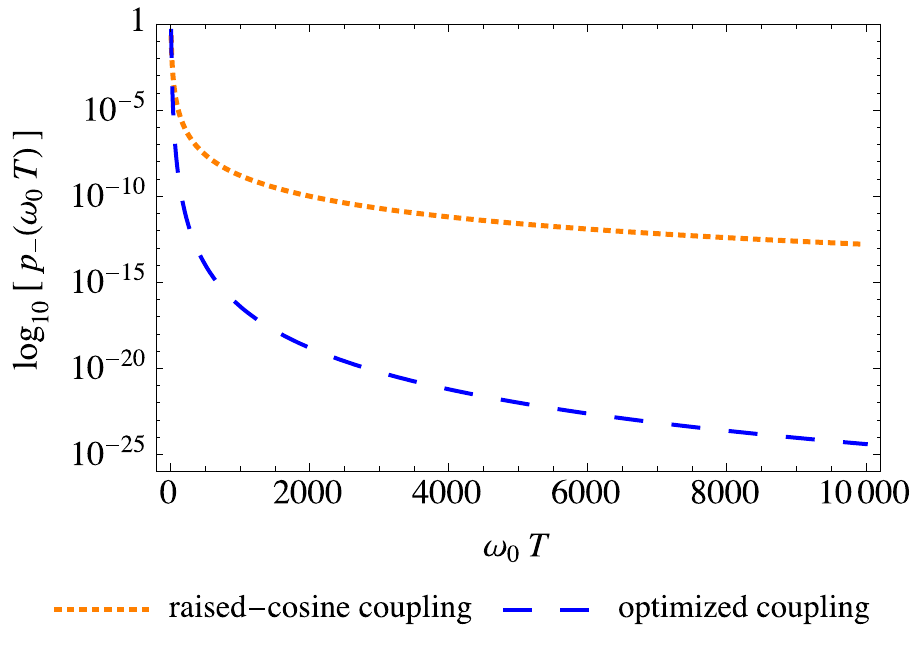}

\caption{\label{fig:rc}(Color online) (a) Raised-cosine function defined by Eq.~\eqref{eq:jfkhjkvhjkvhjkv11881} and the ``optimized'' coupling function defined by Eq.~\eqref{eq:bvdhkjbvd11}. Both coupling functions describe a gradual turn-on and turnoff of the measurement field. The horizontal axis is in units of the measurement time $T$ and the vertical axis is in units of $1/T$. (b) Corresponding reduction of state disturbance. We plot the ratio $p_-=P_-^{(1)}/P_-^\text{const}$ of the transition probability $P_-^{(1)}$ for each coupling to the transition probability $P_-^\text{const}$ for constant coupling, shown as a function of the dimensionless parameter $\omega_0 T$.}

\end{figure}

To illustrate the reduction of the state disturbance obtained for time-dependent couplings, we consider two examples. The raised-cosine function was already studied, more generally, in Ref.~\cite{Schlosshauer:2014:pm} and is defined by
\begin{equation}\label{eq:jfkhjkvhjkvhjkv11881}
g(t)=\frac{1}{T}\left[ 1+\cos\left(\frac{2\pi (t-T/2)}{T}\right)\right] \qquad \text{for $0 \le t \le T$},
\end{equation}
and $g(t)=0$ otherwise [Fig.~\ref{fig:rc}(a)]. From Eq.~\eqref{eq:8dhj7gr7ss82}, the first-order transition amplitude is
\begin{align}\label{eq:8dhj7gr7ss82aaa}
A_-^{(1)}(q,\gamma,T) &= \I \frac{\mu\beta q}{\hbar}\sin\gamma \,\E^{\I \eta} \,\frac{\mathrm{sinc}(\omega_0 T/2)}{1-(\omega_0 T/ 2\pi)^2}.
\end{align}
In a protective measurement, $\omega_0 T$ is chosen to be large to minimize the state disturbance (since the sinc function decays as the inverse of its argument). In this case the damping factor is well approximated by $-(\omega_0 T/ 2\pi)^{-2}$. Disregarding as usual the oscillations represented by the sinc function, Eq.~\eqref{eq:8dhj7gr7ss82aaa} becomes
\begin{align}\label{eq:8dhj7g117gr82as78978}
A_-^{(1)}(q,\gamma,T) &\approx - \I \frac{\mu\beta q}{\hbar}\sin\gamma \,\E^{\I \eta} \frac{\pi^2}{(\omega_0 T/ 2)^3}.
\end{align}
This expression scales as $1/T^3$, which is to be compared to the $1/T$ scaling for constant coupling [see Fig.~\ref{fig:rc}(b)].

We can further reduce the state disturbance by making the turn-on and turnoff of the raised-cosine function \eqref{eq:jfkhjkvhjkvhjkv11881} even more gradual, 
\begin{align}\label{eq:bvdhkjbvd11}
g(t) &= \frac{1}{T}\biggl[ 1 +\frac{4}{3}\cos\biggl(\frac{2\pi (t-T/2)}{T}\biggr) \notag \\ &\quad +\frac{1}{3}\cos\biggl(\frac{4\pi (t-T/2)}{T}\biggr)\biggr] \qquad \text{for $0 \le t \le T$},
\end{align}
and $g(t)=0$ otherwise [Fig.~\ref{fig:rc}(a)]. For $\omega_0 T \gg 1$ and disregarding the oscillations of the sinc function, we find
\begin{equation}\label{eq:bvdhkjbvd20000}
A_-^{(1)}(q,\gamma,T) \approx -\frac{\mu\beta q}{\hbar}  \,\E^{\I \eta} \sin\gamma \frac{4\pi^4}{(\omega_0 T/2)^5}, 
\end{equation}
which scales as $1/T^{5}$, a significant additional reduction of the state disturbance compared to the $1/T^3$ dependence for the raised-cosine function [see Fig.~\ref{fig:rc}(b)].

The condition $\omega_0T\gg 1$ is easily achieved in an experimental setting. As an example, we take again the Stern--Gerlach experiment with potassium atoms. For $B_0=\unit[1]{T}$, $\omega_0^{-1}$ is on the order of $\unit[10^{-11}]{s}$, which is to be compared to the time $T$ required for the atom to traverse the region over which the measurement field is applied. For a width $d=\unit[0.1]{m}$ of this region and an atomic velocity $v = \unit[450]{m/s}$, we have $T \approx \unit[0.2]{ms}$, which is seven orders of magnitude larger than  $\omega_0^{-1}$.

\section{\label{sec:quant-state-reconstr}Successive versus simultaneous protective measurements}

To reconstruct a quantum state, we need to protectively measure multiple observables. To minimize the total state disturbance, should we measure those observables successively or simultaneously? One might expect that a simultaneous measurement will be superior, because for successive measurements the disturbed state produced by an earlier measurement will become the initial state for a subsequent measurement, thus propagating the error. Here we show that, for our model, this is not so: both methods will result in the same state disturbance. 

State reconstruction requires the protective measurement of three spin directions. Consider the measurement fields $\bvec{B}_1^{(k)}(\bvec{x}) = \frac{1}{T} \beta_k q_k \bvec{n}_k$ for orthogonal directions $\bvec{n}_k$, with $k=1,2,3$ (we assume constant coupling), and also consider the combined field  $\bvec{B}_1(\bvec{x}) = \sum_{k=1}^3 \bvec{B}_1^{(k)}(\bvec{x})$. The first-order transition amplitude is now found to be 
\begin{equation}\label{eq:huhu44}
A_-^{(1)}(\bvec{x},T) = \frac{\I}{\hbar} \mu \left( \sum_{k=1}^3 \beta_k q_k \sin\gamma_k \,\E^{\I \eta_k}\right) \mathrm{sinc}\left( \frac{\omega_0 T}{2}\right),
\end{equation}
where $\gamma_k$ is the angle between $\bvec{n}_k$ and $\bvec{B}_0$, and $\eta_k$ is the azimuthal angle of $\bvec{n}_k$. This is simply a sum over the first-order transition amplitudes~\eqref{eq:8dhj7gr7ss82} evaluated separately for the three measurement fields $\bvec{B}_1^{(k)}(\bvec{x})$. Equation~\eqref{eq:huhu44} is to be compared to the transition amplitude for a measurement procedure consisting of three successive protective measurements. We model this procedure as a single protective measurement of duration $3T$ with interaction Hamiltonian 
\begin{multline}\label{eq:1dvhjaabbdhvbdhjv}
\op{H}_\text{int}(\bvec{x}, t) = \\  \begin{cases} -\frac{1}{T} \mu\beta_k (\bopvecgr{\sigma} \cdot
  \bvec{n}_k) \otimes \op{q}_k, &  (k-1)T \le t \le kT, \\ &  \quad k=1,2,3, \\ 0, & \text{otherwise}. \end{cases}
\end{multline}
The corresponding first-order transition amplitude is 
\begin{align}\label{eq:huhu445}
A_-^{(1)}(\bvec{x},T) &= \frac{\I}{\hbar} \mu \left( \sum_{k=1}^3\beta_k q_k  \sin\gamma_k \,\E^{\I \eta_k}\,\E^{\I (k-2) \omega_0 T}\right) \notag \\ &\qquad \times \,\mathrm{sinc}\left( \frac{\omega_0 T}{2}\right),
\end{align}
which, apart from the relative phase factors $\E^{\I (k-2) \omega_0 T}$, is the same as Eq.~\eqref{eq:huhu44}. These phase factors describe rapid oscillations as a function of $T$ and can therefore be considered experimentally irrelevant in the same way as we have disregarded the oscillations of the sinc function. Thus, we have confirmed that successive and simultaneous protective measurements introduce identical amounts of state disturbance in our model.

\section{\label{sec:pointer-shift}Momentum-shift reversals}

Finally, let us explore a hitherto overlooked caveat of protective measurement: even if no state disturbance occurs, the measured direction of the momentum shift may be reversed, resulting in a reconstructed state that can differ drastically from the initial state. Consider the amplitude $A_+(\bvec{x},T) = \braket{+}{\psi(\bvec{x}, T)}$ for the system to be found in the initial state $\ket{+}$ at the conclusion of the measurement. Assuming constant coupling, Eq.~\eqref{eq:vihdgs7cf6gv} gives
\begin{align}\label{eq:sy658hhgvfvihdgs7cf6gv}
A_+(\bvec{x},T) &= \frac{1+\cos\theta(\bvec{x})}{2} \exp\left( \frac{\I \mu B(\bvec{x}) T}{\hbar}\right) \notag \\ &\quad +\frac{1-\cos\theta(\bvec{x})}{2} \exp\left(- \frac{\I \mu B(\bvec{x}) T}{\hbar}\right).
\end{align}
Using $B(\bvec{x}) \approx B_0 + \frac{\beta q}{T}  {\bra{+} \bopvecgr{\sigma} \cdot \bvec{n}  \ket{+}}$ [see Eq.~\eqref{eq:hbvdvbhj}], Eq.~\eqref{eq:sy658hhgvfvihdgs7cf6gv} shows that the state $\ket{+}$ in the final state vector is associated with a superposition of opposite momentum shifts $\pm \Delta \bvec{p} = \pm  \mu\beta  \bra{+} \bopvecgr{\sigma} \cdot \bvec{n}  \ket{+} \bvec{n}$. Only the state corresponding to $+\Delta \bvec{p}$, however, represents the correct pointer shift. From Eq.~\eqref{eq:sy658hhgvfvihdgs7cf6gv}, and using that 
\begin{align}
\cos \theta (\bvec{x}) &= \frac{\bvec{B}(\bvec{x}) \cdot \bvec{e}_z}{B(\bvec{x})} = \frac{1 + \xi  \cos\gamma}{ \sqrt{1 + \xi^2 + 2 \xi \cos\gamma}},
\end{align}
the probability for a measurement of the particle's momentum shift to yield the incorrect value $-\Delta \bvec{p}$ conditional on the system's being found in the state $\ket{+}$ is, to lowest order in $\xi$, 
\begin{align}\label{eq:djdjzaaa}
P (\xi, \gamma) \approx \left(\frac{1}{2}\xi\sin\gamma\right)^4.
\end{align}
Because this probability is of order $O(\xi^4)$, it is typically negligible. The fact that the probability amplitude corresponding to Eq.~\eqref{eq:djdjzaaa} is of order $1/T^2$ explains why it was missed not only in the zeroth-order limit of infinitely large $T$, but also in the first-order treatment of Refs.~\cite{Schlosshauer:2014:pm,Schlosshauer:2014:tp}. 

Even though the probability is typically small, we may still ask what the consequence of measuring $-\Delta \bvec{p}$ instead of $+\Delta \bvec{p}$ would be, as far as state reconstruction is concerned. The reversed momentum shift translates to the expectation value of $\bopvecgr{\sigma} \cdot \bvec{n}$ in the orthogonal state $\ket{-}$. Suppose that one of three successive protective measurements of observables $\bopvecgr{\sigma} \cdot \bvec{n}_i$ has resulted in a reversed momentum shift, and take $\bopvecgr{\sigma} \cdot \bvec{n}_3$ to be this failed measurement. Then the expectation value of $\bopvecgr{\sigma} \cdot \bvec{n}_3$ in the initial state $\ket{+}$ determined from the momentum shift would be $-\cos\gamma$, where $\gamma$ is the angle between $\bvec{n}_3$ and $\bvec{B}_0$. In the $\{\ket{+},\ket{-}\}$ basis, the corresponding density matrix reconstructed from the three protective measurements of $\bopvecgr{\sigma} \cdot \bvec{n}_i$ would then be
\begin{align}\label{eq:djd7g8ygt8j}
\op{\rho}= \begin{pmatrix} \sin^2\gamma & -\I\sin\gamma\cos\gamma\,\E^{-\I\eta} \\ \I\sin\gamma\cos\gamma\,\E^{\I\eta} & \cos^2\gamma \end{pmatrix},
\end{align}
rather than $\op{\rho}= \ketbra{+}{+}=\left(\begin{smallmatrix} 1&0\\0&0 \end{smallmatrix}\right)$, where $\eta$ is the azimuthal angle of $\bvec{n}_3$. The pure state corresponding to Eq.~\eqref{eq:djd7g8ygt8j} is $\ket{\psi}=\sin\gamma\ket{+}+\I \E^{\I\eta}\cos\gamma\ket{-}$. For example, if $\gamma=45^\circ$, then $\ket{\psi}=2^{-1/2} \left(\ket{+}+\I \E^{\I\eta}\ket{-}\right)$, an equal-weight superposition of $\ket{+}$ and $\ket{-}$. 
The functional dependence of the fidelity $F(\op{\rho},\ket{+})=\sqrt{\bra{+}\op{\rho}\ket{+}}=\sin \gamma$ may be understood as follows. If $\gamma=0$, then the failed measurement is in the direction of the quantization axis of the initial spin state $\ket{+}$ and indicates the expectation value $-1$ associated with the state $\ket{-}$, while the expectation values of the protective measurements in the two orthogonal directions are zero. Thus the conjunction of these three measurements would lead one to conclude that the system's state must be $\ket{-}$, and therefore the fidelity will be zero. Conversely, if $\gamma=90^\circ$, then the expectation value of the protective measurement along this direction is zero, and thus a sign flip of this expectation value leaves the fidelity unaffected.

\section{\label{sec:disc-concl}Discussion and conclusions}

Physically realizable protective measurements inevitably disturb the initial state of the system, implying a nonzero probability for the measurement to fail. Fundamentally, this disturbance is rooted in the tradeoff between quantum-state disturbance and information gain in a quantum measurement \cite{Fuchs:1996:op}, as well as in the fact that the maximum possible information gain does not depend on the particular implementation of the quantum measurement \cite{Ariano:1996:om}. To determine the likelihood of success of a protective measurement and the fidelity of quantum-state reconstruction based on protective measurements, one needs to be able to quantify the state disturbance, as well as the faithfulness of the measurement outcome. 

In this paper we have analyzed these issues in the context of a concrete model that may be experimentally realizable using a setup of the Stern--Gerlach type. While this model has been used previously \cite{Aharonov:1993:qa, Dass:1999:az} to illustrate basic features of protective measurement, the essential issues studied in this paper had not yet been considered. One of our main results is that if the strength of the weak inhomogeneous magnetic field producing the measurement interaction does not vary in time during the measurement interval, then the amount of disturbance of the initial state is completely quantified by two parameters. The first parameter is the ratio between the measurement field and the strong uniform magnetic field providing the protection of the initial state. The second parameter is the angle between the unknown direction of the protection field and the experimentally chosen direction of the measurement field. We found that the transition probability shows, to good approximation, a quadratic dependence on the field-strength parameter and a sinusoidal dependence on the angle parameter. Thus, weakening the measurement field reduces the state disturbance, despite the fact that the measurement time $T$ is simultaneously increased by the same factor (this is so because the measurement field is inversely proportional to $T$). The increase of the state disturbance as a function of the angle parameter can be understood from the complementarity principle, since increasing the angle corresponds to measuring a spin component in a direction further away from the direction represented by the quantization axis of the initial spin state. 

The state disturbance can be reduced not only by decreasing the relative strength of the measurement field, but also by turning the field on and off in a gradual fashion, as was already shown more generally in Ref.~\cite{Schlosshauer:2014:pm}. Our results illustrate that such a gradual turn-on and turnoff accomplishes a much more effective reduction of the state disturbance than could be achieved by merely making the measurement field weaker. Although it is in principle easy to realize any desired smooth time dependence of the measurement field by gradually changing the current in the electromagnet, the experimental challenge lies in appropriately timing the field such that it is gradually turned on just as the particle enters the measurement region. This also requires that only a single particle traverses the measurement field at any given time. Among other measures, reaching the single-particle regime may require a suitably narrow collimating slit for minimizing the particle flux issuing from the source. 

For a protective measurement to be successful, it needs to not only leave the system in its initial state at the conclusion of the measurement, but the shift of the apparatus pointer must also be a faithful representation of the expectation value of the measured observable in this initial state. Our analysis shows that the issues of state disturbance and faithful pointer shift (here realized as a momentum transfer) are distinct and require individual attention. In particular, we found that even when the system is left in its initial state at the end of the measurement interaction, the measurement may still result in the wrong pointer shift (albeit with a probability proportional to $1/T^4$ that is likely to be negligibly small in practice). This pointer shift corresponds to a change in the particle's momentum in the opposite direction from the direction associated with the expectation value of the measured spin component in the initial quantum state. As we have shown, such an error can have severe consequences for the fidelity of the quantum-state reconstruction.

If one uses a measurement field that is inhomogeneous in all three directions in space, one can correspondingly impart a momentum shift with three distinct spatial components. Measuring these components provides the same information as gained from successive protective measurements using three different (nonphysical) measurement fields that are inhomogeneous in only one direction. We showed that the resulting state disturbance is also the same, in agreement with what would one expect from the general relationship between information gain and disturbance in a quantum measurement. This suggests a more general result concerning protective measurements, namely, that multiple successive measurements are equivalent, both in terms of the resulting pointer shifts and the cumulative state disturbance, to carrying out the same measurements simultaneously. Nonetheless, in a concrete experimental situation one of these two possible measurement procedures may be easier to realize. For example, in the spin measurement considered in this paper, a simultaneous implementation using a measurement field that is inhomogeneous in all three directions in space is not only the physically relevant case (since the field must be divergence-free), but also makes it unnecessary to arrange three separate Stern--Gerlach apparatuses.

Despite several promising theoretical proposals \cite{Aharonov:1993:jm,Anandan:1993:uu,Nussinov:1998:yy,Dass:1999:az}, the experimental realization of protective measurements is still an open challenge. Our analysis indicates that an implementation of the measurement scheme studied in this paper may be within the parameter regime of existing Stern--Gerlach experiments, such as those based on a beam of evaporated potassium atoms. Both the dimensions of the apparatus and the field inhomogeneities typically found in such experiments are suited for producing an appreciable, macroscopic pointer shift, although meeting the weak-measurement condition will require a sufficiently strong uniform magnetic field in the vicinity of \unit[1--10]{T}. While a more careful estimate may need to include consideration of experimental imperfections and other factors, our aim here was to focus on the issue of state disturbance and to show how it constrains the experimental parameters. While the experimental challenges are considerable, our analysis suggests that an implementation of the measurement scheme studied here may well be feasible in the near future. 


\begin{thebibliography}{15}%
\makeatletter
\providecommand \@ifxundefined [1]{%
 \@ifx{#1\undefined}
}%
\providecommand \@ifnum [1]{%
 \ifnum #1\expandafter \@firstoftwo
 \else \expandafter \@secondoftwo
 \fi
}%
\providecommand \@ifx [1]{%
 \ifx #1\expandafter \@firstoftwo
 \else \expandafter \@secondoftwo
 \fi
}%
\providecommand \natexlab [1]{#1}%
\providecommand \enquote  [1]{``#1''}%
\providecommand \bibnamefont  [1]{#1}%
\providecommand \bibfnamefont [1]{#1}%
\providecommand \citenamefont [1]{#1}%
\providecommand \href@noop [0]{\@secondoftwo}%
\providecommand \href [0]{\begingroup \@sanitize@url \@href}%
\providecommand \@href[1]{\@@startlink{#1}\@@href}%
\providecommand \@@href[1]{\endgroup#1\@@endlink}%
\providecommand \@sanitize@url [0]{\catcode `\\12\catcode `\$12\catcode
  `\&12\catcode `\#12\catcode `\^12\catcode `\_12\catcode `\%12\relax}%
\providecommand \@@startlink[1]{}%
\providecommand \@@endlink[0]{}%
\providecommand \url  [0]{\begingroup\@sanitize@url \@url }%
\providecommand \@url [1]{\endgroup\@href {#1}{\urlprefix }}%
\providecommand \urlprefix  [0]{URL }%
\providecommand \Eprint [0]{\href }%
\providecommand \doibase [0]{http://dx.doi.org/}%
\providecommand \selectlanguage [0]{\@gobble}%
\providecommand \bibinfo  [0]{\@secondoftwo}%
\providecommand \bibfield  [0]{\@secondoftwo}%
\providecommand \translation [1]{[#1]}%
\providecommand \BibitemOpen [0]{}%
\providecommand \bibitemStop [0]{}%
\providecommand \bibitemNoStop [0]{.\EOS\space}%
\providecommand \EOS [0]{\spacefactor3000\relax}%
\providecommand \BibitemShut  [1]{\csname bibitem#1\endcsname}%
\let\auto@bib@innerbib\@empty
\bibitem [{\citenamefont {Aharonov}\ and\ \citenamefont
  {Vaidman}(1993)}]{Aharonov:1993:qa}%
  \BibitemOpen
  \bibfield  {author} {\bibinfo {author} {\bibfnamefont {Y.}~\bibnamefont
  {Aharonov}}\ and\ \bibinfo {author} {\bibfnamefont {L.}~\bibnamefont
  {Vaidman}},\ }\href@noop {} {\bibfield  {journal} {\bibinfo  {journal} {Phys.
  Lett. A}\ }\textbf {\bibinfo {volume} {178}},\ \bibinfo {pages} {38}
  (\bibinfo {year} {1993})}\BibitemShut {NoStop}%
\bibitem [{\citenamefont {Aharonov}\ \emph {et~al.}(1993)\citenamefont
  {Aharonov}, \citenamefont {Anandan},\ and\ \citenamefont
  {Vaidman}}]{Aharonov:1993:jm}%
  \BibitemOpen
  \bibfield  {author} {\bibinfo {author} {\bibfnamefont {Y.}~\bibnamefont
  {Aharonov}}, \bibinfo {author} {\bibfnamefont {J.}~\bibnamefont {Anandan}}, \
  and\ \bibinfo {author} {\bibfnamefont {L.}~\bibnamefont {Vaidman}},\
  }\href@noop {} {\bibfield  {journal} {\bibinfo  {journal} {Phys. Rev. A}\
  }\textbf {\bibinfo {volume} {47}},\ \bibinfo {pages} {4616} (\bibinfo {year}
  {1993})}\BibitemShut {NoStop}%
\bibitem [{\citenamefont {Anandan}\ and\ \citenamefont
  {Vaidman}(1996)}]{Aharonov:1996:fp}%
  \BibitemOpen
  \bibfield  {author} {\bibinfo {author} {\bibfnamefont {Y.~A.~J.}\
  \bibnamefont {Anandan}}\ and\ \bibinfo {author} {\bibfnamefont
  {L.}~\bibnamefont {Vaidman}},\ }\href@noop {} {\bibfield  {journal} {\bibinfo
   {journal} {Found. Phys.}\ }\textbf {\bibinfo {volume} {26}},\ \bibinfo
  {pages} {117} (\bibinfo {year} {1996})}\BibitemShut {NoStop}%
\bibitem [{\citenamefont {{Hari Dass}}\ and\ \citenamefont
  {Qureshi}(1999)}]{Dass:1999:az}%
  \BibitemOpen
  \bibfield  {author} {\bibinfo {author} {\bibfnamefont {N.~D.}\ \bibnamefont
  {{Hari Dass}}}\ and\ \bibinfo {author} {\bibfnamefont {T.}~\bibnamefont
  {Qureshi}},\ }\href@noop {} {\bibfield  {journal} {\bibinfo  {journal} {Phys.
  Rev. A}\ }\textbf {\bibinfo {volume} {59}},\ \bibinfo {pages} {2590}
  (\bibinfo {year} {1999})}\BibitemShut {NoStop}%
\bibitem [{\citenamefont {Vaidman}(2009)}]{Vaidman:2009:po}%
  \BibitemOpen
  \bibfield  {author} {\bibinfo {author} {\bibfnamefont {L.}~\bibnamefont
  {Vaidman}},\ }in\ \href@noop {} {\emph {\bibinfo {booktitle} {Compendium of
  Quantum Physics: Concepts, Experiments, History and Philosophy}}},\ \bibinfo
  {editor} {edited by\ \bibinfo {editor} {\bibfnamefont {D.}~\bibnamefont
  {Greenberger}}, \bibinfo {editor} {\bibfnamefont {K.}~\bibnamefont
  {Hentschel}}, \ and\ \bibinfo {editor} {\bibfnamefont {F.}~\bibnamefont
  {Weinert}}}\ (\bibinfo  {publisher} {Springer},\ \bibinfo {address}
  {Berlin/Heidelberg},\ \bibinfo {year} {2009})\ pp.\ \bibinfo {pages}
  {505--508}\BibitemShut {NoStop}%
\bibitem [{\citenamefont {Gao}(2014)}]{Gao:2014:cu}%
  \BibitemOpen
  \bibinfo {editor} {\bibfnamefont {S.}~\bibnamefont {Gao}},\ ed.,\ \href@noop
  {} {\emph {\bibinfo {title} {Protective Measurement and Quantum Reality:
  Towards a New Understanding of Quantum Mechanics}}}\ (\bibinfo  {publisher}
  {Cambridge University Press, Cambridge},\ \bibinfo {year} {2014})\BibitemShut {NoStop}%
\bibitem [{\citenamefont {Schlosshauer}(2014)}]{Schlosshauer:2014:pm}%
  \BibitemOpen
  \bibfield  {author} {\bibinfo {author} {\bibfnamefont {M.}~\bibnamefont
  {Schlosshauer}},\ }\href@noop {} {\bibfield  {journal} {\bibinfo  {journal}
  {Phys. Rev. A}\ }\textbf {\bibinfo {volume} {90}},\ \bibinfo {pages} {052106}
  (\bibinfo {year} {2014})}\BibitemShut {NoStop}%
\bibitem [{Note1()}]{Note1}%
  \BibitemOpen
  \bibinfo {note} {It is important to emphasize that although we have specified
  the direction of the field $\protect \ensuremath {\protect \mathbf {B}}_0$ to
  enable the subsequent mathematical description of the model, in an actual
  realization of the protective measurement the field's direction and magnitude
  will be \protect \emph {a priori} unknown (see also the discussion in
  Refs.~\cite {Aharonov:1993:jm} and \cite {Dass:1999:az}). Indeed, an observer
  who knows $\protect \ensuremath {\protect \mathbf {B}}_0$ would also know
  $\protect \mathaccentV {hat}05E{H}_S$ and could therefore perform a simple
  projective measurement in the eigenbasis of $\protect \mathaccentV
  {hat}05E{H}_S$ to determine the spin state, eliminating the need to perform a
  protective measurement.}\BibitemShut {Stop}%
\bibitem [{Note2()}]{Note2}%
  \BibitemOpen
  \bibinfo {note} {As already noted in Ref.~\cite {Aharonov:1993:jm}, since
  Eq.~\protect \textup {\hbox {\mathsurround \z@ \protect \normalfont
  (\ignorespaces \ref {eq:measfield}\unskip \@@italiccorr )}} has nonzero
  divergence, it violates Maxwell's equations and cannot represent a real
  physical magnetic field. However, a suitable divergence-free inhomogeneous
  field is easily constructed \cite {Anandan:1993:uu}. Such a field effectively
  acts as a superposition of three fields of the kind given by Eq.~\protect
  \textup {\hbox {\mathsurround \z@ \protect \normalfont (\ignorespaces \ref
  {eq:measfield}\unskip \@@italiccorr )}} and leads to the same cumulative
  momentum shift and state disturbance (see also Sec.~\ref
  {sec:quant-state-reconstr}). Without loss of generality, we may therefore
  restrict our attention to the field defined by Eq.~\protect \textup {\hbox
  {\mathsurround \z@ \protect \normalfont (\ignorespaces \ref
  {eq:measfield}\unskip \@@italiccorr )}}.}\BibitemShut {Stop}%
\bibitem [{\citenamefont {Schlosshauer}\ and\ \citenamefont
  {Claringbold}(2014)}]{Schlosshauer:2014:tp}%
  \BibitemOpen
  \bibfield  {author} {\bibinfo {author} {\bibfnamefont {M.}~\bibnamefont
  {Schlosshauer}}\ and\ \bibinfo {author} {\bibfnamefont {T.~V.~B.}\
  \bibnamefont {Claringbold}},\ }in\ \href@noop {} {\emph {\bibinfo {booktitle}
  {Protective Measurement and Quantum Reality: Towards a New Understanding of
  Quantum Mechanics}}},\ \bibinfo {editor} {edited by\ \bibinfo {editor}
  {\bibfnamefont {S.}~\bibnamefont {Gao}}}\ (\bibinfo  {publisher} {Cambridge
  University Press, Cambridge},\ \bibinfo {year} {2014}), pp.~180--194\BibitemShut {NoStop}%
\bibitem [{\citenamefont {Daybell}(1967)}]{Daybell:1967:sg}%
  \BibitemOpen
  \bibfield  {author} {\bibinfo {author} {\bibfnamefont {M.~D.}\ \bibnamefont
  {Daybell}},\ }\href@noop {} {\bibfield  {journal} {\bibinfo  {journal} {Am.
  J. Phys.}\ }\textbf {\bibinfo {volume} {35}},\ \bibinfo {pages} {637}
  (\bibinfo {year} {1967})}\BibitemShut {NoStop}%
\bibitem [{\citenamefont {Fuchs}\ and\ \citenamefont
  {Peres}(1996)}]{Fuchs:1996:op}%
  \BibitemOpen
  \bibfield  {author} {\bibinfo {author} {\bibfnamefont {C.~A.}\ \bibnamefont
  {Fuchs}}\ and\ \bibinfo {author} {\bibfnamefont {A.}~\bibnamefont {Peres}},\
  }\href@noop {} {\bibfield  {journal} {\bibinfo  {journal} {Phys. Rev. A}\
  }\textbf {\bibinfo {volume} {53}},\ \bibinfo {pages} {2038} (\bibinfo {year}
  {1996})}\BibitemShut {NoStop}%
\bibitem [{\citenamefont {D'Ariano}\ and\ \citenamefont
  {Yuen}(1996)}]{Ariano:1996:om}%
  \BibitemOpen
  \bibfield  {author} {\bibinfo {author} {\bibfnamefont {G.~M.}\ \bibnamefont
  {D'Ariano}}\ and\ \bibinfo {author} {\bibfnamefont {H.~P.}\ \bibnamefont
  {Yuen}},\ }\href@noop {} {\bibfield  {journal} {\bibinfo  {journal} {Phys.
  Rev. Lett.}\ }\textbf {\bibinfo {volume} {76}},\ \bibinfo {pages} {2832}
  (\bibinfo {year} {1996})}\BibitemShut {NoStop}%
\bibitem [{\citenamefont {Anandan}(1993)}]{Anandan:1993:uu}%
  \BibitemOpen
  \bibfield  {author} {\bibinfo {author} {\bibfnamefont {J.}~\bibnamefont
  {Anandan}},\ }\href@noop {} {\bibfield  {journal} {\bibinfo  {journal}
  {Found. Phys. Lett.}\ }\textbf {\bibinfo {volume} {6}},\ \bibinfo {pages}
  {503} (\bibinfo {year} {1993})}\BibitemShut {NoStop}%
\bibitem [{\citenamefont {Nussinov}(1998)}]{Nussinov:1998:yy}%
  \BibitemOpen
  \bibfield  {author} {\bibinfo {author} {\bibfnamefont {S.}~\bibnamefont
  {Nussinov}},\ }\href@noop {} {\bibfield  {journal} {\bibinfo  {journal}
  {Found. Phys.}\ }\textbf {\bibinfo {volume} {28}},\ \bibinfo {pages} {865}
  (\bibinfo {year} {1998})}\BibitemShut {NoStop}%
\end{thebibliography}

%

\end{document}